\documentclass{article}

\usepackage{arxiv}

\usepackage[utf8]{inputenc} % allow utf-8 input
\usepackage[T1]{fontenc}    % use 8-bit T1 fonts
\usepackage[super,sort&compress,comma]{natbib} 
\usepackage{amsmath}
\usepackage{mhchem}
\usepackage{graphicx}
\usepackage{hyperref}       % hyperlinks
\usepackage{url}            % simple URL typesetting
\usepackage{booktabs}       % professional-quality tables
\usepackage{amsfonts}       % blackboard math symbols
\usepackage{nicefrac}       % compact symbols for 1/2, etc.
\usepackage{microtype}      % microtypography
\usepackage{lipsum}
\usepackage{amssymb}
\usepackage{bm}
\usepackage{gensymb}
\usepackage[format=plain,justification=justified,singlelinecheck=false,font={stretch=1.125,small},labelfont=bf,labelsep=space]{caption}

\title{A first-principles machine-learning force field for heterogeneous ice nucleation on microcline feldspar}

\author{
  Pablo M. Piaggi \\
  Department of Chemistry, Princeton University, Princeton, NJ 08544, USA \\
  \texttt{ppiaggi@princeton.edu}
  \And
  Annabella Selloni \\
  Department of Chemistry, Princeton University, Princeton, NJ 08544, USA \\
  \And
  Athanassios Z. Panagiotopoulos \\
  Department of Chemical and Biological Engineering, Princeton University, Princeton, NJ 08544, USA \\
  \And
  Roberto Car \\
  Department of Chemistry, Princeton University, Princeton, NJ 08544, USA \\
  Department of Physics, Princeton University, Princeton, NJ 08544, USA\\
  \And
  Pablo G. Debenedetti \\
  Department of Chemical and Biological Engineering, Princeton University, Princeton, NJ 08544, USA \\
}

\begin{document}
\maketitle

\begin{abstract}
The formation of ice in the atmosphere affects precipitation and cloud properties, and plays a key role in the climate of our planet. Although ice can form directly from liquid water at deeply supercooled conditions, the presence of foreign particles can aid ice formation at much warmer temperatures. Over the past decade, experiments have highlighted the remarkable efficiency of feldspar minerals as ice nuclei compared to other particles present in the atmosphere. However, the exact mechanism of ice formation on feldspar surfaces has yet to be fully understood.
Here, we develop a first-principles machine-learning model for the potential energy surface aimed at studying ice nucleation at microcline feldspar surfaces.
The model is able to reproduce with high-fidelity the energies and forces derived from density-functional theory (DFT) based on the SCAN exchange and correlation functional.
Our training set includes configurations of bulk supercooled water, hexagonal and cubic ice, microcline, and fully-hydroxylated feldspar surfaces exposed to vacuum, liquid water, and ice.
We apply the machine-learning force field to study different fully-hydroxylated terminations of the (100), (010), and (001) surfaces of microcline exposed to vacuum.
Our calculations suggest that terminations that do not minimize the number of broken bonds are preferred in vacuum.
We also study the structure of supercooled liquid water in contact with microcline surfaces, and find that water density correlations extend up to around 10 \AA~from the surfaces.
Finally, we show that the force field maintains a high accuracy during the simulation of ice formation at microcline surfaces, even for large systems of around 30,000 atoms.
Future work will be directed towards the calculation of nucleation free energy barriers and rates using the force field developed herein, and understanding the role of different microcline surfaces on ice nucleation.
\end{abstract}

\section{Introduction}

Anthropogenic climate change and the associated extreme weather events are currently underscoring, perhaps more than at any other point in human history, the need to better understand the different factors that contribute to the climate and weather in our planet\cite{masson2021climate,stott2016climate}.
One such factor is the formation of ice in the atmosphere, which affects cloud microphysics, radiative properties, and precipitation formation\cite{pruppacher1980microphysics,Vergara18}.
The initial stage in the formation of ice is an activated process called ice nucleation, which occurs below the freezing temperature.
In this process, order and density fluctuations in liquid water give rise to the fleeting appearance of microscopic ice clusters.
From a thermodynamic point of view, this process can be rationalized as consisting of two competing forces.
On the one hand, the formation of an ice cluster creates a liquid water-ice interface with an associated free energy cost.
On the other hand, below the freezing temperature ice has a lower chemical potential than liquid water and thus there is a bulk thermodynamic driving force favoring its formation.
The competition between a thermodynamically favourable bulk contribution and an unfavourable interfacial contribution results in a free energy barrier for a given \emph{critical cluster} size.
The system must therefore surmount this free energy barrier in order to proceed with the liquid water to ice transformation.
Decreasing the temperature results in a greater driving force for the formation of ice and a concomitant reduction in the free energy barrier.
In the absence of foreign particles, \emph{homogeneous} ice nucleation takes place at around 38 K below freezing at ambient pressure.
At these conditions the free energy barrier is sufficiently low and leads to a rapid formation of ice in micrometer-sized water droplets.
However, nucleation can also occur at higher temperatures aided by ice nucleating particles (INP), which are able to lower the free energy barrier.
The latter process is called \emph{heterogeneous} ice nucleation.

A large variety of aerosol particles are known to act as INP in the atmosphere, including mineral dusts, soot, and biological particles\cite{kanji2017overview,hoose2012heterogeneous,wilson2015marine}.
Among different minerals present in aerosol dust particles, feldspars have been shown to be highly active for ice nucleation\cite{atkinson2013importance}, and are also the most common type of mineral in Earth's crust\cite{klein2013earth}.
Feldspars are tectosilicates, and span a broad range of crystal structures and chemical compositions, which include K, Na, and/or Ca\cite{klein2013earth}.
In the last decade, a significant body of experimental evidence has been produced that points to K-rich feldspar as the most important type of INP\cite{atkinson2013importance,yakobi2013feldspar,augustin2014immersion,zolles2015identification,harrison2016not,peckhaus2016comparative,welti2019ice,harrison2019ice}, capable of nucleating ice at relatively low supercoolings of about 20 K.
It has been suggested by Kiselev \textit{et al.}~that ice is formed at the high-energy ($100$) surfaces of K-rich feldspar exposed at defects, and that the prismatic ($10\bar{1}0$) plane of ice is aligned with the ($100$) plane of feldspar\cite{kiselev2017active}.
Although the $(100)_\mathrm{feldspar}/(10\bar{1}0)_\mathrm{ice\:Ih}$ orientation relationship was originally observed in deposition mode\cite{kiselev2017active} (heterogeneous nucleation from water vapor), it was recently observed as well in immersion mode (heterogeneous nucleation from liquid water) \cite{keinert2022mechanism}.
Static molecular simulations reported by Kiselev \textit{et al.}~provided evidence in favor of the observed orientation relationship\cite{kiselev2017active}.
However, subsequent molecular dynamics simulations by Patey and coworkers failed to show high ice-nucleation activity of the $(100)$, $(010)$, and $(001)$ surfaces of K-feldspar\cite{soni2019simulations,kumar2021molecular}.
It can be inferred from the latter studies that the surface feature responsible for ice nucleation has not been fully identified, that the atomistic structure of the interfaces is not well known, or that the models for the interatomic interactions used in molecular simulations fail to capture important physical processes.
Indeed, all molecular simulations reported above relied on empirical models for the interatomic interactions, which are based on relatively simple functional forms and contain a few parameters fit to experimental properties.

Here, we develop a first-principles machine-learning (ML) force field for the study of ice nucleation on microcline feldspar (\ce{KAlSi3O8}).
The ML model reproduces the potential energy surface (PES) derived from quantum-mechanical density-functional theory (DFT) calculations \cite{Hohenberg64,Kohn65}.
It is thus able to describe complex changes in the interatomic interactions modulated by the environment, including bond forming/breaking and polarization.
We obtain a highly-accurate model for liquid water and ice in contact with the fully-hydroxylated $(100)$, $(010)$, and $(001)$ surfaces of microcline.
Below, we describe the procedure to construct the force field, and benchmark the properties of water, ice, and microcline.
We also study the surface terminations of microcline, and the structure of water at the surfaces.
Finally, we test the reliability of the force field for the simulation of ice nucleation at microcline/water interfaces.

\begin{figure}[t]
\centering
  \includegraphics[width=\textwidth]{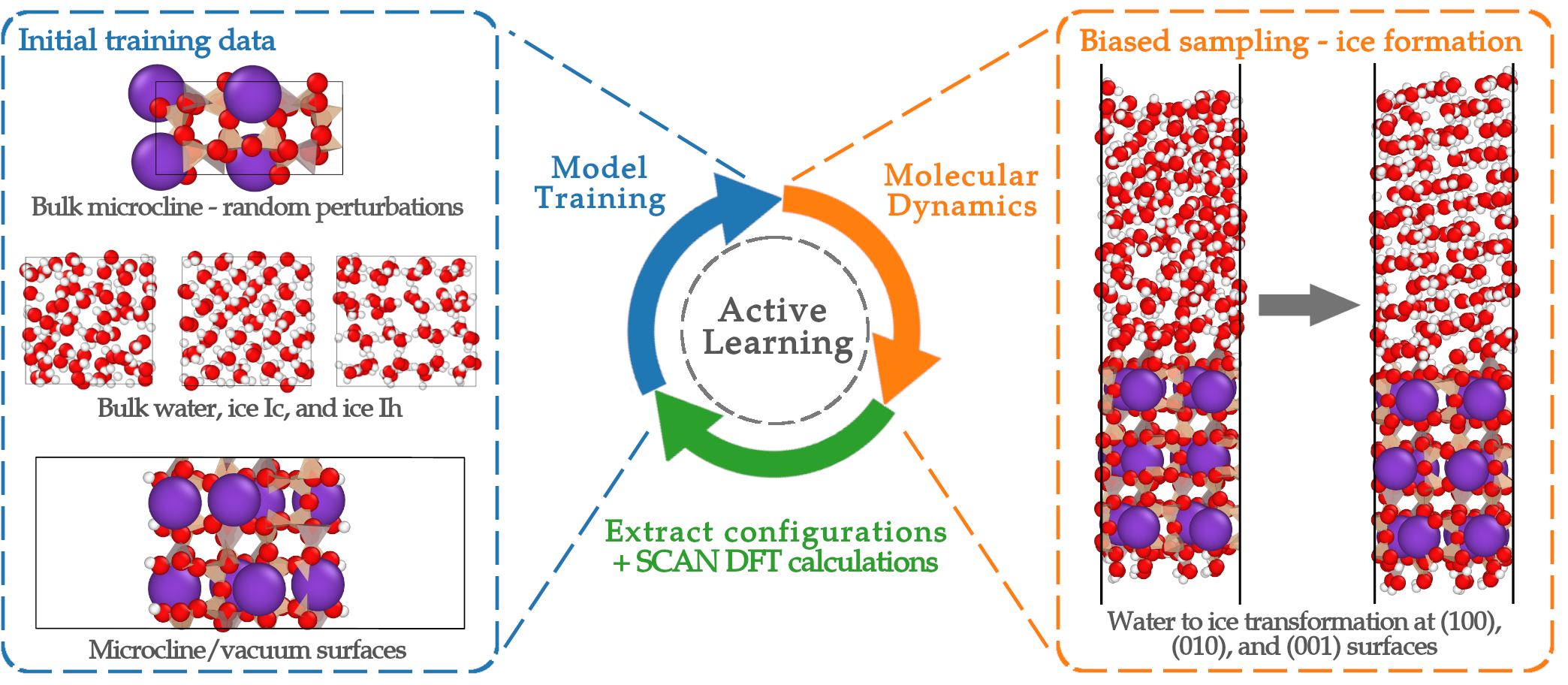}
  \caption{Schematic summary of the training procedure for the ML force field based on active learning. See text for more information about the training process. In the atomistic configurations \ce{K}, \ce{O}, and \ce{H} atoms are shown as purple, red, and white spheres, respectively, with their corresponding van der Waals radii. \ce{Si} and \ce{Al} atoms are depicted using brown and gray tetrahedra, respectively, with oxygens located at their corners. Images obtained with Ovito\cite{Stukowski09}.}
  \label{fig:training_summary}
\end{figure}

\section{Model training}

In the last few years, ML models for the PES trained on energies and forces derived from quantum-mechanical electronic-structure theory have become a standard tool to perform first principles molecular dynamics simulations at an affordable computational cost.
While direct \textit{ab initio} molecular dynamics (AIMD) simulations are limited to system sizes of a few hundreds of atoms and total simulation times of tens of ps, ML force fields can provide access to system sizes and simulation times several orders of magnitude greater.

In this work, we constructed one such model using the Deep Potential Molecular Dynamics (DeePMD) methodology\cite{Zhang18,Zhang18end}, with an eye towards simulating ice nucleation on microcline from first principles.
A schematic summary of the training process is shown in Fig.~\ref{fig:training_summary}.
Briefly, the procedure is based on first training an initial model with a small number of training data including configurations of bulk liquid water, ice, microcline feldspar, and feldspar/vacuum interfaces with their corresponding \textit{ab initio} energies and forces.
The energies and forces were obtained from DFT calculations based on the Strongly Constrained and Appropriately Normed (SCAN) functional\cite{Sun15,Sun16}.
SCAN has been extensively benchmarked for the study of water and ice\cite{Chen17,Gartner20,Piaggi21,Zhang21}, and provides a good balance between accuracy and computational cost.
Afterwards, the model was iteratively improved using an active learning procedure\cite{zhang2019active,smith2018less,podryabinkin2019accelerating,schran2021machine} based on performing molecular dynamics simulations of ice formation at the feldspar/water surfaces using the model trained in the previous iteration, extracting configurations in which the error in the forces is high, computing energies and forces for such configurations using DFT, incorporating the configurations to the training set, and training a new model (see Fig.~\ref{fig:training_summary}).
The repetition of these steps until a convergence criterion is met leads to a model that accurately reproduces the phenomenon under study, in this case, ice nucleation at microcline surfaces.
Because ice nucleation is a rare event, it typically cannot be observed spontaneously during a standard molecular dynamics simulation.
For this reason, we used enhanced sampling simulations\cite{Valsson16review} to drive the formation of ice at microcline surfaces in relatively short 1 ns simulations.

The final training set obtained via active learning contained around $\sim$11,000 snapshots in total, including configurations of bulk liquid water, ice I$_\mathrm{h}$, ice I$_\mathrm{c}$, and microcline feldspar, as well as feldspar/vacuum, feldspar/water, feldspar/ice, and water/vacuum interfaces.
The training data on feldspar interfaces included configurations with exposed $(100)$, $(010)$, and $(001)$ crystallographic surfaces.
The training process was carried out using the DeePMD-kit\cite{Wang18}, and simulations were performed with the molecular dynamics engine LAMMPS\cite{thompson2022lammps} augmented by the DeePMD-kit and the enhanced sampling plugin PLUMED\cite{Tribello14,Bonomi19}. 
Further information about the training set, the training process, and other simulations details are given  in the ESI.

The model presented here is a significant step forward in the realistic description of this type of system for several reasons.
First, the ML force field is fully predictive because it does not rely on any empirical information.
Second, it reproduces the full PES derived from quantum-mechanical electronic-structure calculations and 
thus it is sensitive to subtle changes in the electronic structure arising from changes in the chemical environments.
Lastly, the force field is polarizable, it has a many-body functional form (at variance with simpler pair potentials), and can describe bond formation and cleavage.
On the other hand, the ML force field developed in this work also has several limitations.
First, the accuracy of the underlying PES on which the model is trained is limited by the choice of approximate functional in the reference DFT calculations.
Here we chose the SCAN functional which does not reproduce quantitatively some properties of water (see results below and refs.~\citenum{Chen17,Gartner20,Piaggi21,Zhang21}).
Second, we perform classical dynamics in which nuclear quantum effects, typically important in systems containing hydrogen, are neglected. 
Finally, the ML force field is short-ranged with a 6 \AA~cutoff and therefore neglects long-range interactions present in the DFT calculations (see ref.~\citenum{zhang2022deep} for a possible methodology to include this type of interaction).

\section{Results}

\subsection{Accuracy of the model}

\begin{figure}[t]
\centering
  \includegraphics[width=\textwidth]{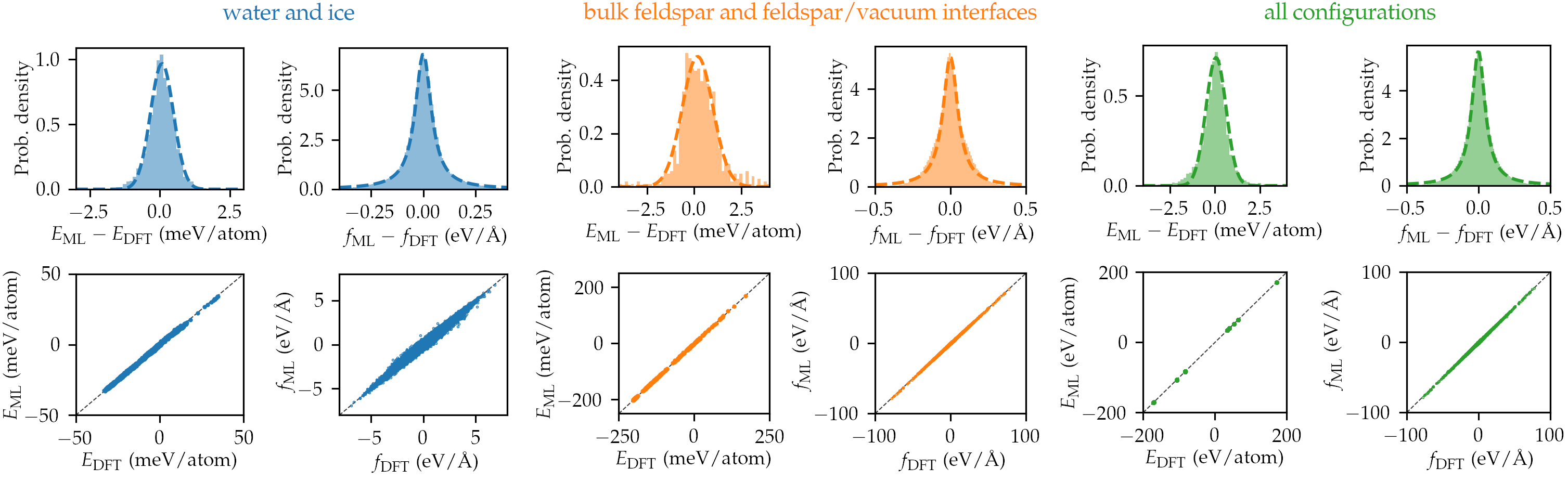}
  \caption{Analysis of the accuracy of the trained model for liquid water and ice configurations (blue), bulk feldspar and feldspar/vacuum interfaces (orange), and the whole training set (green). We show the distribution of errors in the energy $E_\mathrm{ML}-E_\mathrm{DFT}$ and the forces $f_\mathrm{ML}-f_\mathrm{DFT}$. The errors in energy and forces were fitted to a Gaussian and Lorentzian probability density function, respectively, and the corresponding curves are shown with a dashed line. Parity plots $E_\mathrm{ML}$ vs. $E_\mathrm{DFT}$ and $f_\mathrm{ML}$ vs. $f_\mathrm{DFT}$ are also shown.}
  \label{fig:model_accuracy}
\end{figure}

We analyze the accuracy of the ML model of the \textit{ab initio} PES obtained in this work using the root mean square (RMS) errors,
\begin{align}
    \epsilon^\mathrm{RMS}_E = & \sqrt{\left \langle \left( \frac{E_\mathrm{ML}(\mathbf{R})-E_\mathrm{DFT}(\mathbf{R})}{N}\right)^2\right \rangle} \\
    \epsilon^\mathrm{RMS}_f = & \sqrt{\left \langle \frac{1}{3N} \left \| \mathbf{f}_\mathrm{ML}(\mathbf{R})-\mathbf{f}_\mathrm{DFT}(\mathbf{R})\right \| ^2\right \rangle}
\end{align}
where $\epsilon^\mathrm{RMS}_E$ and $\epsilon^\mathrm{RMS}_f$ are the RMS errors in the energy and the forces, respectively, $E_\mathrm{ML}(\mathbf{R})$ and $E_\mathrm{DFT}(\mathbf{R})$ are the potential energies calculated by the ML model and using DFT for a configuration with $N$ atoms and atomic coordinates $\mathbf{R}$, $\mathbf{f}_\mathrm{ML}(\mathbf{R})$ and $\mathbf{f}_\mathrm{DFT}(\mathbf{R})$ are the $3N$-dimensional forces calculated by the ML model and using DFT, and $\langle \cdot \rangle$ denotes an average over the configurations in all or a subset of the training set.
In the ESI we show the training curves for the final model.
%As seen in Fig.~\ref{fig:model_training},
These curves show that $\epsilon^\mathrm{RMS}_E$ and $\epsilon^\mathrm{RMS}_f$ calculated over the batches (single configurations) used for training reach an average value of around 0.5 meV/atom and 100 meV/\AA, respectively, by the end of the training process.
Once the training process was completed, we evaluated errors over the entire training set of 11,172 configurations and found RMS errors in the energy and forces of $\epsilon^\mathrm{RMS}_E=0.81$ meV/atom and $\epsilon^\mathrm{RMS}_f=106$ meV/\AA, respectively.
Such errors are of the same order of magnitude as those of other state-of-the-art machine learning models for the potential energy surface\cite{Deringer20,Zhang21}.
We also evaluated the RMS errors over two subsets of the training set.
Considering only configurations of liquid water and ice, the RMS errors in the energy and forces were $\epsilon^\mathrm{RMS}_E=0.45$ meV/atom and $\epsilon^\mathrm{RMS}_f=100$ meV/\AA, respectively.
Instead, taking into account configurations of bulk feldspar and feldspar/vacuum interfaces, the RMS errors in the energy and forces were $\epsilon^\mathrm{RMS}_E=1.10$ meV/atom and $\epsilon^\mathrm{RMS}_f=112$ meV/\AA, respectively.
This analysis shows that the model describes with approximately uniform accuracy the very different chemical environments in our training set.

Now we turn to analyze the distribution of the errors in the energy and forces.
In Fig.~\ref{fig:model_accuracy} we show the distribution of the errors $E_\mathrm{ML}-E_\mathrm{DFT}$ and $f_\mathrm{ML}-f_\mathrm{DFT}$ for all configurations in the training set, for liquid water and ice, and for bulk feldspar and feldspar/vacuum interfaces.
The distribution of errors in the energy and forces follow approximately a Gaussian and Lorentzian distribution, respectively, centered at zero and with spreads comparable to the RMS values reported above.
Furthermore, the parity plots $E_\mathrm{ML}$ vs. $E_\mathrm{DFT}$ and $f_\mathrm{ML}$ vs. $f_\mathrm{DFT}$ also show excellent correlation between the training data and the model predictions across all configurations in the training set.

Having established the accuracy of the ML model to reproduce the reference DFT energies and forces, we now move to characterize the properties of water, ice, microcline, and their interfaces in the next few sections. 

\subsection{Properties of bulk microcline feldspar}

We consider the properties of bulk microcline with stoichiometry \ce{KAlSi3O8} and without \ce{Al}/\ce{Si} disorder.
The zero-temperature lattice constants of bulk microcline were calculated via energy minimization with respect to atomic coordinates and cell dimensions using direct DFT calculations and the ML model.
We also computed the average lattice constants in MD simulations at 300 K and 1 bar driven by the ML model.
The results are shown in Table \ref{tbl:microcline_lattice} and are also compared to experimental lattice constants.
The agreement between calculated and experimental lattice constants is very good, and the average error with respect to experiments\cite{bailey1969refinement,kroll1987determining} is around 0.5 \%.

\begin{table}[t]
\small
\begin{center}
  \caption{Lattice constants of microcline feldspar. Experimental values at room temperature are compared with the prediction of the SCAN DFT functional at 0 K, and with the ML model developed here at 0 K and 300 K. All experimental and simulated lattice constants correspond to ambient pressure conditions (1 bar). The percent error of SCAN DFT with respect to experiments\cite{kroll1987determining} (Exp.$^b$) is also shown.}
  \label{tbl:microcline_lattice}
  \begin{tabular*}{0.9\textwidth}{@{\extracolsep{\fill}}lllllll}
    \hline
      & Exp.$^a$ & Exp.$^b$ & SCAN DFT (0 K) & SCAN-ML (0 K) & SCAN-ML (300 K) & Error DFT (\%) \\
    \hline
 a (\AA) & 8.5784 & 8.592 & 8.58 & 8.55 & 8.58 & -0.14 \\
 b (\AA) & 12.96	& 12.963 & 13.00 & 13.02 & 13.02 & 0.29 \\
 c (\AA) & 7.2112 & 7.222 & 7.25 & 7.22 & 7.22 & 0.39 \\
 $\alpha$ (\degree) & 90.3 & 90.62	& 91.15 & 90.68 & 90.59 & 0.58 \\
 $\beta$ (\degree) & 116.03 & 115.9 & 116.23 & 116.21 & 116.09 & 0.28 \\ 
 $\gamma$ (\degree) &  89.125 & 87.6 & 87.49 & 87.55 & 87.59 & -0.13 \\
    \hline
  \end{tabular*}
  \end{center}
 $^a$ Ref.~\citenum{bailey1969refinement},
 $^b$ Ref.~\citenum{kroll1987determining}
\end{table}

\subsection{Properties of water and ice}

We evaluated the ability of the ML model to reproduce a variety of properties of the bulk phases of water at 1 bar.
Many properties of water and ice in an ML model based on SCAN DFT have previously been calculated by us\cite{Piaggi21} and here we reevaluate some properties for the model developed in this work.
Even though we use the same functional (SCAN), the inclusion of additional atom types leads to a completely different ML model, so this test is necessary. 
We calculated the melting temperature $T_m$ of ice I$_\mathrm{h}$ as described in ref.~\citenum{Piaggi21}, the enthalpy of fusion $\Delta H_f$, and the density change upon melting $(\rho_{l}-\rho_{Ih})/\rho_{Ih}$ (\%).
The values for these quantities for the current model, and the experimental counterparts, are summarized in Table \ref{tbl:water_ice_properties}.
\begin{table}[b]
\small
\begin{center}
  \caption{Properties of liquid water and ice I$_\mathrm{h}$. Melting temperature $T_m$ of ice I$_\mathrm{h}$, enthalpy of fusion $\Delta H_f$, density change upon melting $(\rho_{l}-\rho_{Ih})/\rho_{Ih}$ (\%), and temperature of maximum liquid water density ($T_\mathrm{MD}$) with respect to $T_m$}
  \label{tbl:water_ice_properties}
  \begin{tabular*}{\textwidth}{@{\extracolsep{\fill}}lllll}
    \hline
      & $T_m$ (K) & $\Delta H_f$ (k$_\mathrm{B}$ T) & $(\rho_{l}-\rho_{Ih})/\rho_{Ih}$ (\%) & $T_\mathrm{MD}-T_m$ (K) \\
    \hline
 SCAN ML (this work) & 307(2) & 3.00(1) & 9 & $\sim$10\\
 SCAN (Ref.~\citenum{Piaggi21}) & 308(2) & 2.9(2) & 6 & $\sim$15\\
 Experiment (Ref.~\citenum{NISTWebBook01}) & 273.15 & 2.65 & 9 & 4\\
    \hline
  \end{tabular*}
  \end{center}
\end{table}
Consistent with previous calculations\cite{Piaggi21} we find a significant overestimation of the melting temperature of 34 K, that we attribute to an artificial strengthening\cite{sharkas2020self} of the hydrogen bonds by the SCAN functional.
The experimental density change upon melting of 9\% is captured with high accuracy by the model.

\begin{figure}[t]
\centering
  \includegraphics[width=0.8\textwidth]{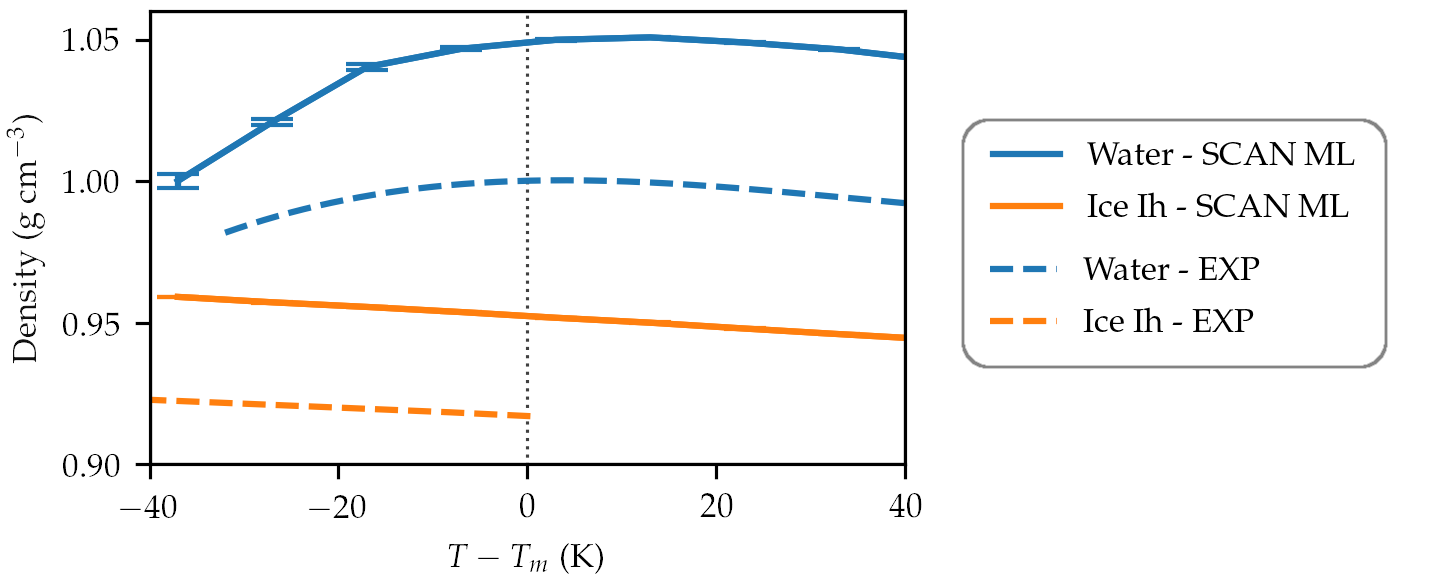}
  \caption{Density of liquid water and ice I$_\mathrm{h}$ calculated using the SCAN ML model developed in this work. Error bars are shown at each simulated temperature and a linear interpolation of the data points is shown with lines. For several data points the error bars are too small to be observed. We also include experimental results for comparison\cite{Chase86,Hare87,Rottger94}.}
  \label{fig:water_density}
\end{figure}

We also evaluated the density of liquid water and hexagonal ice as a function of temperature, and we present the results in Fig.~\ref{fig:water_density}.
The model correctly predicts that ice floats on liquid water, as found with direct AIMD using the SCAN functional\cite{Chen17}.
However, the density of both phases is overestimated by around 5\% across all studied temperatures.
Furthermore, the density of liquid water has a maximum around 10 K above the melting temperature, in good agreement with the well-known density maximum of real water at 4 K above freezing at 1 bar.
These results exhibit the same trends found in ref.~\citenum{Piaggi21} for another ML model based on the SCAN functional.
Although the model presented here has several limitations in the quantitative description of the properties of water and ice, the qualitative trends are captured very well.
We refer the reader to ref.~\citenum{Piaggi21} for an in-depth discussion of the properties of SCAN water and ice.

\subsection{Stability of feldspar/vacuum interfaces}

\begin{figure}[t]
\centering
  \includegraphics[width=0.95\textwidth]{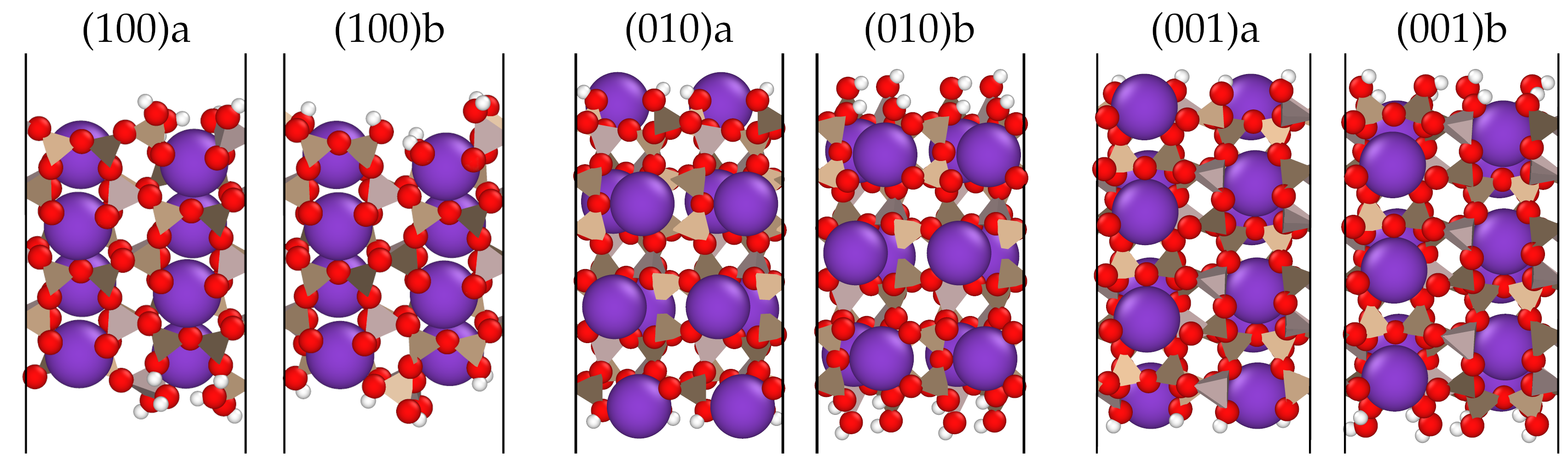}
  \caption{Different fully-hydroxylated terminations of the $(100)$, $(010)$ and $(001)$ surfaces of microcline feldspar. See main text for further information. Details of the visualization of the atomistic configurations are the same as in Fig.~\ref{fig:training_summary}.}
  \label{fig:surfaces_confs}
\end{figure}

The detailed surface structure can impact significantly the formation of ice.
For this reason we evaluated the stability of several terminations of the $(100)$, $(010)$, and $(001)$ surfaces of microcline exposed to vacuum.
The choice of studied crystallographic planes is motivated by the fact that microcline can be experimentally cleaved at the $(010)$ and $(001)$ planes\cite{kiselev2017active,keinert2022mechanism}.
We have also studied the $(100)$ surface guided by the insight put forward by Kiselev et al.\ that ice may nucleate at these surfaces\cite{kiselev2017active}. 
The studied surfaces are depicted in Fig.~\ref{fig:surfaces_confs}.
As in previous studies\cite{kiselev2017active,soni2019simulations,pedevilla2016can}, we assumed full hydroxylation of the surfaces and we studied only terminations that can be generated by the addition of dissociated water molecules (\ce{H} + \ce{OH}).
For the $(010)$ and $(001)$ surfaces we studied two terminations that correspond to cuts along planes of minimum bond breakage (termination type a) and cuts along planes that do not minimize the number of broken bonds (termination type b).
Cuts along planes of minimum bond breakage are good candidates for the most stable terminations, in the absence of hydroxylation.
Indeed, refs.~\citenum{soni2019simulations} and \citenum{pedevilla2016can} have simulated interfaces resulting from cuts along planes of minimum bond breakage (termination type a).

In order to evaluate the stability of the different terminations we calculated the zero-temperature surface energy for slabs of different thicknesses using the ML force field developed in this work.
The calculated surface energy in the limit of infinitely-thick slabs\cite{fiorentini1996extracting} is shown in Table \ref{tbl:surface_energy}.
\begin{table}[t]
\small
\begin{center}
  \caption{Zero-temperature surface energy of different fully-hydroxylated terminations of the $(100)$, $(010)$, and $(001)$ surfaces of microcline feldspar exposed to vacuum. The results labeled SCAN-ML employ the ML potential developed in this work to fully relax the atomic coordinates and to calculate the potential energy of feldspar slabs. The results labelled SCAN-DFT use direct DFT calculations to recompute the potential energy after relaxing the structure with the ML potential. The methodology for the calculation of the surface energy and the errors is described in the ESI.}
  \label{tbl:surface_energy}
  \begin{tabular*}{0.9\textwidth}{@{\extracolsep{\fill}}lllllll}
    \hline
      &  $(100)$a & $(100)$b & $(010)$a & $(010)$b & $(001)$a & $(001)$b  \\
    \hline
    SCAN-ML (mJ m$^-2$) & 111(14) & 205(1) & 283(3) & 16(7) & 140(4) & 17(2) \\
    SCAN-DFT (mJ m$^-2$) & 125(7) & 186(4) & 255(9) & 47(16) & 146(1) & 66(8) \\
    \hline
  \end{tabular*}
  \end{center}
\end{table}
According to our calculations, the surface energy of the type b terminations is lower than that of type a terminations both for the $(010)$ and the $(001)$ surfaces exposed to vacuum.
Thus, the most stable terminations in vacuum are of type b, i.e., correspond to cuts that do not minimize the number of broken bonds in the microcline structure.
We argue that type b terminations are stabilized due to a large number of dissociated water molecules which can form covalent bonds with the surface.
Furthermore, type a terminations expose \ce{K+} ions to vacuum, which may be unfavourable.
As far as we know, the experimental surface structure of microcline feldspar has not been elucidated yet.
However, the surface structure of orthoclase, a higher temperature polymorph closely related to microcline, has been studied by high-resolution x-ray reflectivity\cite{fenter2000atomic,fenter2003structure}.
The electronic density profiles of orthoclase $(010)$ and $(001)$ surfaces in contact with water obtained in that work are compatible with the type a terminations.
It should be noted, however, that the degree of surface hydroxylation is not accessible to such studies.

The choice of termination for the $(100)$ surface poses a greater challenge than for the $(010)$ and $(001)$ cases, due to the crystallographic complexity of the surface.
We have evaluated the stability of the $(100)$ surface termination proposed in ref.~\citenum{kiselev2017active} (termination type a) and a related one that we propose in this work (termination type b).
Both terminations have the same number of dissociated water molecules at the surface.
These terminations are shown in Fig.~\ref{fig:surfaces_confs}.
For the $(100)$ surface, our calculations shown in Table \ref{tbl:surface_energy} predict the type a termination to be more stable than the type b termination (surface energy lower for type a termination).
Type a termination was studied in refs.~\citenum{kiselev2017active} and \citenum{soni2019simulations}.
Experimental information about the structure of the $(100)$ surface is not available due to the difficulties in obtaining microcline or orthoclase samples that expose this plane.

Comparison of the most stable termination for each of the studied surfaces leads to the conclusion that surfaces with the $(010)$ and the $(001)$ crystallographic planes exposed to vacuum have a similar stability, while the ($100$) surface has a lower stability.
This trend is in agreement with the experimentally observed surfaces\cite{kiselev2017active,keinert2022mechanism}.
We also validated the results of the ML force field by recalculating the energies using direct DFT calculations with the SCAN functional.
The results of these calculations, reported in Table \ref{tbl:surface_energy}, show the same trends reported above for the ML force field.

\subsection{Atomistic structure of large feldspar/water interfaces}

The computational efficiency of the ML model developed here allows us to simulate relatively large feldspar/water interfaces.
\begin{figure}[b]
\centering
  \includegraphics[width=0.95\textwidth]{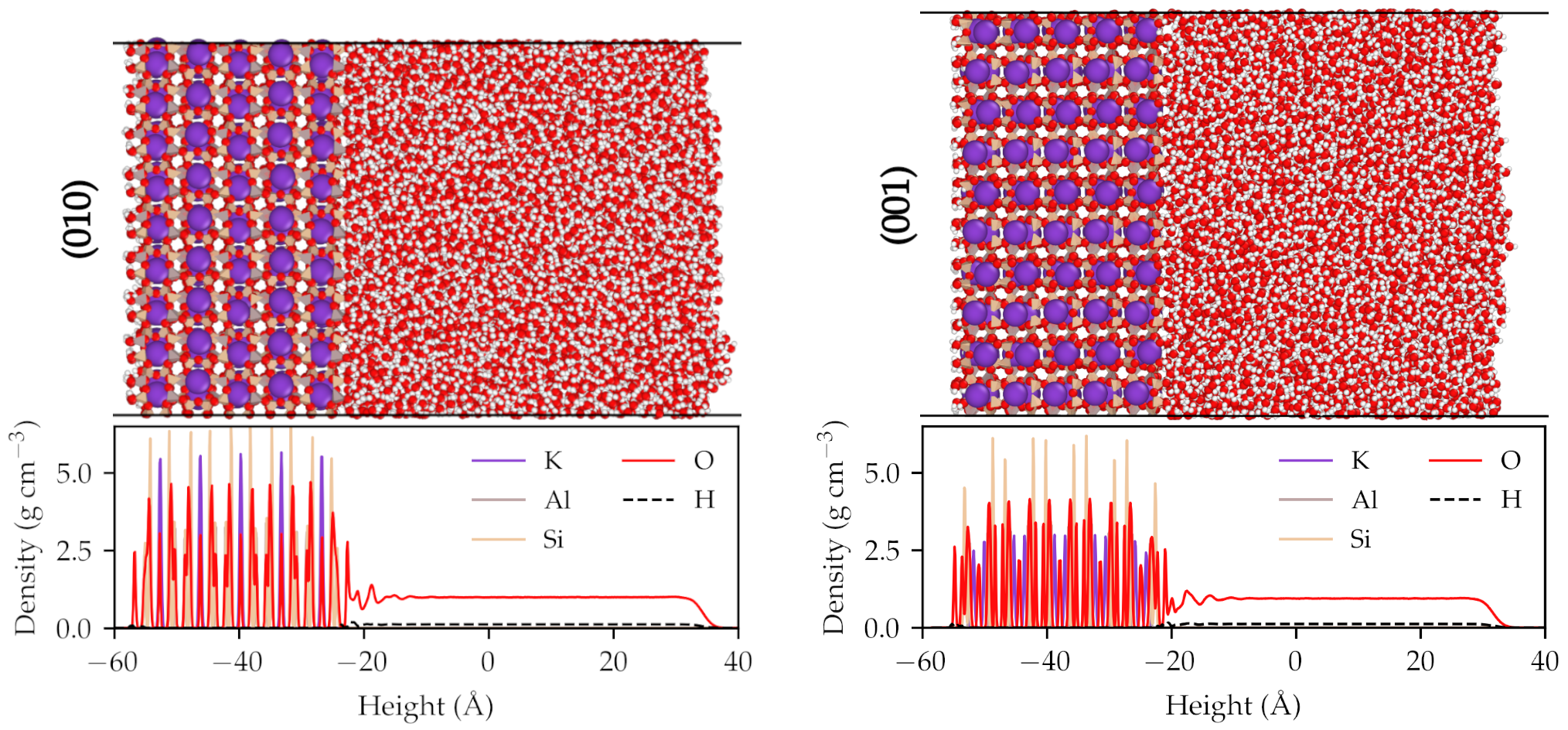}
  \caption{Structure of water at surfaces with the $(010)$ and the $(001)$ crystallographic planes of microcline feldspar exposed to liquid water. Details of the visualization of the atomistic configurations are the same as in Fig.~\ref{fig:training_summary}. The density profiles vs.~height with respect to the surface for each chemical species were averaged over 4-ns simulations at constant temperature 290 K (17 K of supercooling) using configurations extracted every 1 ps.}
  \label{fig:surface_water_structure}
\end{figure}
In Fig.~\ref{fig:surface_water_structure} we show microcline feldspar slabs exposing one fully-hyroxylated b-type surface to liquid water, and the other identical surface to vacuum.
We restrict the analysis to type b terminations of the $(010)$ and the $(001)$ surfaces, because preliminary tests suggested that type a terminations would require a force field trained to describe the exchange of \ce{K+} ions between the surface and water.
The configurations shown in Fig.~\ref{fig:surface_water_structure} for the $(010)$ and the $(001)$ surfaces contain $\sim$30,000 atoms each, a system size well out of reach for direct AIMD simulations.
We performed 4-ns simulations at constant temperature of 290 K (17 K of supercooling relative to  $T_m$ predicted by SCAN-ML) and we calculated the average density of each atomic species (\ce{K}, \ce{Al}, \ce{Si}, \ce{O}, and \ce{H}) as a function of the height with respect to the surface.
The computed density profiles for the $(010)$ and the $(001)$ surfaces are shown in Fig.~\ref{fig:surface_water_structure}.
In both cases, the surface induces structuring of water, observed as density variations with respect to the bulk water density, that extend up to a distance of around 10 \AA~from the surface.
The maximum distance of such correlations found here is larger than the structuring distance of 5 \AA~reported by Fenter et al.\cite{fenter2003structure} for water at orthoclase surfaces.
The density profiles reported here correspond to slabs terminated at planes that do not minimize the number of broken bonds (type b), instead of slabs terminated at planes of minimum bond breakage (type a) as studied by Fenter et al\cite{fenter2003structure}.

Another relevant feature of the surface is the degree of surface protonation.
We controlled the coordination number of oxygen atoms with protons, and found that the number of oxygens with coordination equal to one (non-bridging oxygens) remained constant throughout the simulation.
This indicates that the $(010)$ and the $(001)$ surfaces were fully hydroxylated during the entire simulations.
We remark, however, that the ML model was not specifically trained to reproduce surface deprotonation and the barriers for proton transfer may not be captured accurately in this model.
Studying surface deprotonation with greater detail is an important subject for future work.

In order to test the reliability of the dynamics generated by the ML model, during the simulation we estimated the errors in the forces using four independently-trained ML models as proposed in ref.~\citenum{zhang2019active} (see the ESI for a definition of the error based on an ensemble of ML models).
The average and maximum error in the forces were 0.03~eV/\AA~and 0.2 eV/\AA, respectively, indicating that the dynamics generated by the ML force field reproduce with high fidelity the DFT potential energy surface.
Note also that these configurations have three different interfaces, namely, vacuum/feldspar, feldspar/water, and water/vacuum.
In spite of this complexity, the ML model is reliable for driving the dynamics of such configurations over long timescales.

We also tested whether ice forms spontaneously at these surface at lower temperatures.
We performed simulations at 270 K and 280 K, corresponding to supercoolings of 27 K and 37 K, for the $(010)$ and $(001)$ surfaces.
Neither surface was able to trigger ice nucleation, even after 50 ns of simulation time.
Furthermore, the dynamics at these temperatures were also stable and the errors in the forces were comparable to the errors reported above at 290 K.

\subsection{Ice clusters at the feldspar/water interfaces}

In the previous section, we have shown the ability of the ML force field developed in this work to describe feldspar/water interfaces.
Now we turn our attention to the formation of ice at the feldspar/water interface.
Since ice does not form spontaneously at the surface during a standard molecular dynamics simulation, we introduced a bias potential that drives the transformation of water into ice.
The bias potential is a function of an order parameter, which we choose to be the $Q_6$ Steinhardt parameter\cite{Steinhardt83} frequently employed for crystal nucleation studies\cite{vanDuijneveldt92,tenWolde96}.
The order parameter is defined using only oxygen atoms, and a weighting function is used to focus the effect of the bias potential within a 15 \AA~sphere centered at the feldspar/water interface. 
The bias potential is chosen to be harmonic with a center that evolves linearly from a $Q_6$ value compatible with liquid water to a $Q_6$ value compatible with ice, over a 10-ns simulation.
For further details about the biasing protocol we refer the reader to the ESI.

\begin{figure}[t]
\centering
  \includegraphics[width=0.8\textwidth]{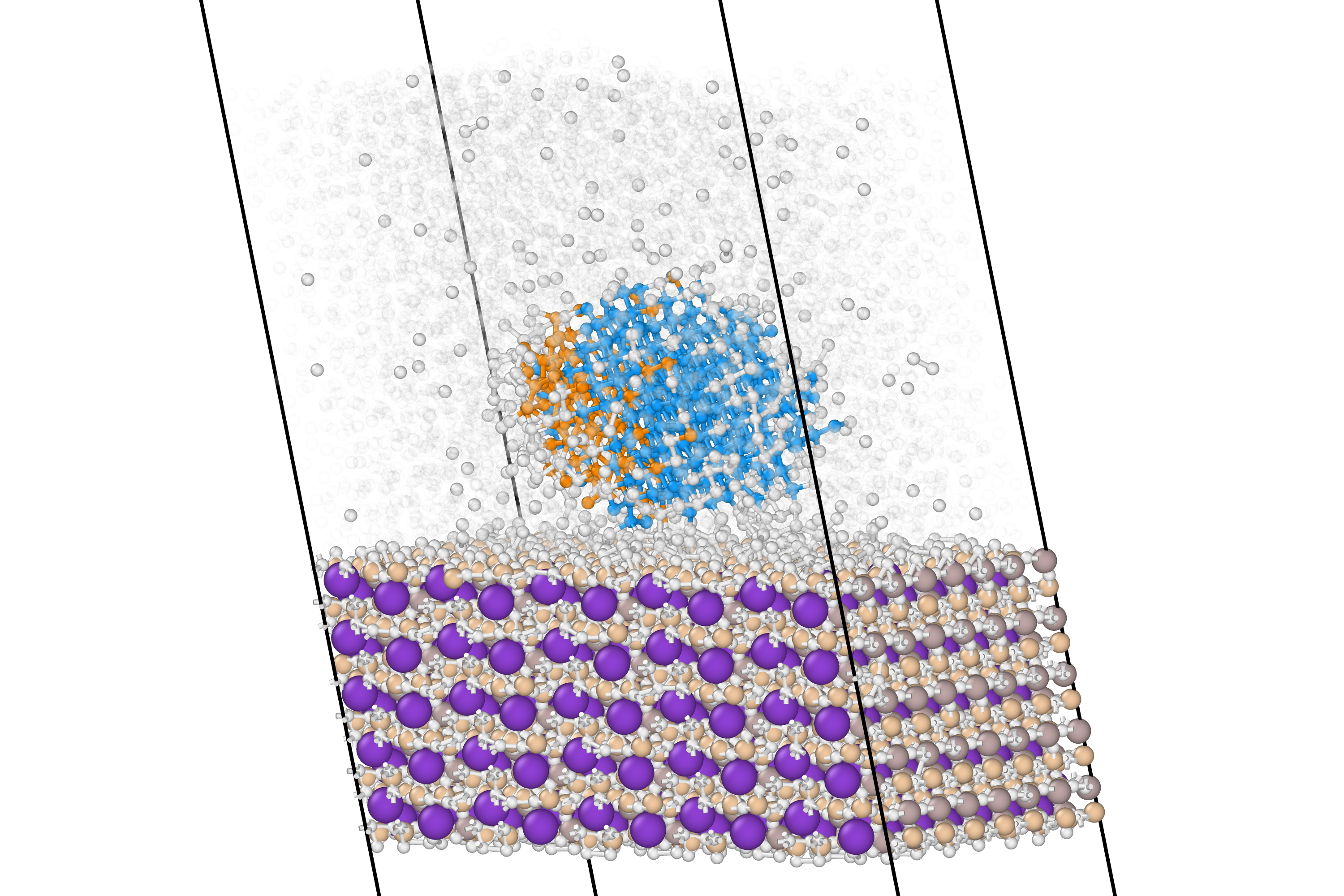}
  \caption{Ice cluster at the $(001)$ surface of microcline in contact with water at 280 K (27 K of supercooling). \ce{K}, \ce{Al}, and \ce{Si} atoms are shown as purple, gray, and brown spheres, respectively, with their corresponding van der Waals radii. \ce{O} atoms with environments compatible with ice I$_\mathrm{h}$ and I$_\mathrm{c}$ are shown in orange and blue, respectively. \ce{O} atoms with other environments are shown as white spheres. In order to easily visualize the solid cluster, we display with high transparency \ce{O} atoms with displacement magnitudes larger than 3 \AA~over the previous 1 ns of simulation. For clarity, \ce{H} atoms are not shown. Bonds are displayed between \ce{O} atoms with interatomic distances smaller than 3 \AA.}
  \label{fig:cluster}
\end{figure}

During the biased simulation of the $(001)$ surface of feldspar in contact with water at 280 K, we observe the formation of an ice cluster at the surface (shown in Fig.~\ref{fig:cluster}).
A structural analysis algorithm\cite{Larsen16} reveals that the ice cluster is mainly formed by ice I$_\mathrm{c}$ yet it also exhibits a region of ice I$_\mathrm{h}$.
The alternating structures of I$_\mathrm{c}$ and I$_\mathrm{h}$ are characteristic of stacking faults in ice.
Furthermore, at the feldspar/water interface the structural analysis fails to identify water molecules as solid-like, but an analysis of atomic displacements shows that the interfacial water has very low-mobility and is thus solid-like.
Another interesting characteristic of ice formed at the $(001)$ microcline surface is the possible orientation relationship between crystallographic planes of ice and microcline.
An analysis of the orientation relationship does not show correspondences between planes of low Miller indices of ice and microcline.

We also analyzed the reliability of the dynamics driven by the ML model during the formation of the ice cluster.
We monitored microcline's surface structure and surface protonation state, and observed that it did not change during the simulation.
Furthermore, the average and maximum errors in the forces remained similar to the errors reported in the previous section.
We thus conclude that the model is well-trained for the study of ice nucleation at microcline surfaces.

\section{Conclusions}

We constructed a first-principles machine-learning model of the potential energy surface, with the purpose of investigating the process of ice nucleation on microcline.
The training data for the model contains five different chemical species (\ce{K}, \ce{Al}, \ce{Si}, \ce{O}, and \ce{H}), four bulk phases (liquid water, ice I$_\mathrm{h}$, ice I$_\mathrm{c}$, microcline), and seven interfaces ($(100)$/water, $(010)$/water, $(010)$/water, $(100)$/vacuum, $(010)$/vacuum, $(010)$/vacuum, and water/vacuum).
We demonstrated that the model reproduces with high-fidelity the reference \textit{ab initio} energies and forces, in spite of the complexity of the system and process under study.

We benchmarked the prediction of several properties of liquid water, ice, and microcline by our force field, and found overall good agreement with the experimental measurements.
Some properties of water and ice, for instance, the melting temperature of ice I$_\mathrm{h}$ and the densities, exhibit the expected qualitative trends yet differ quantitavely by around 5-10 \% with respect to experiments.
The discrepancies are mainly attributed to the limitations of the SCAN DFT functional.

We also analyzed the stability of several surface terminations of the $(100)$, $(010)$, and $(001)$ crystallographic planes of microcline exposed to vacuum.
We restricted the analysis to fully-hydroxylated surfaces.
Our calculations show that the most stable terminations of the $(010)$ and $(001)$ surfaces in vacuum do not correspond to planes of minimum bond breakage.
We interpret this finding as resulting from two different effects.
On the one hand, the formation of a greater number of bonds with the surface \ce{H} and \ce{OH} groups stabilizes the surface, in spite of the larger number of broken bonds of microcline.
On the other hand, exposing \ce{K+} ions to vacuum, as in the terminations of minimum bond breakage, leads to an unfavourable electrostatic contribution.
We note that the stability ranking of these terminations might change if surfaces are exposed to liquid water, as a consequence of the screening of the electric field of the \ce{K+} ions and the formation of hydrogen bonds with the surface \ce{H} and \ce{OH} groups.
Indeed, x-ray reflectivity experiments in orthoclase have found the surface termination of the $(010)$ and $(001)$ surfaces in contact with water to be that of minimum bond breakage (assuming that the surface structure of microcline is similar to that of orthoclase).
Our calculations rely on several assumptions, including the predictive power of the SCAN DFT calculations and full surface hydroxylation.
Further computational and experimental work should be aimed at elucidating the atomistic structures of the $(010)$ and $(001)$ surfaces in contact with water, including the evaluation of the degree of surface hydroxylation.
For the $(100)$ surface, we evaluated the stability of two terminations and our results support that the termination proposed in ref.~\citenum{kiselev2017active} is the most stable in vacuum.

We also tested the ability of our model to simulate relatively large interfacial systems, of around 30,000 atoms, with \textit{ab initio} accuracy.
We showed that the force field developed in this work generates stable and highly-accurate dynamics (average error in forces $\sim$ 30 meV/\AA) for such systems.
We investigated the structure of water at the $(100)$ and $(010)$ microcline surface, and found that the structure of water is affected up to distances of around 10 \AA.
The influence of the surface on the structure of water is more pronounced in our simulations than in experimental measurements in orthoclase\cite{fenter2003structure}, where it was found that the density of water was affected up to distances of 5 \AA.
This discrepancy may be a result of the somewhat different structure of microcline and orthoclase, a different surface termination, and/or a limited resolution in the experimental measurement.

Although ice nucleation was not observed spontaneously during standard molecular dynamics simulations, we drove the formation of ice at the $(001)$ surface using an appropriately designed bias potential.
The dynamics during the transformation of liquid water into ice were also highly-accurate with an average error in the forces of $\sim$ 30 meV/\AA.
The ice cluster thus formed contained stacking faults and did not exhibit an orientation relationship between planes of low Miller indices of ice and microcline.

We are presently working on the determination of nucleation free energy barriers and on characterizing the ice/microcline orientation relationship at the $(100)$, $(010)$, and $(001)$ microcline surfaces.
%The results are deferred to a future publication.

\section*{Author Contributions}
P.M.P performed research and wrote the original draft; all authors conceived the project, designed research, discussed results, and reviewed and edited the manuscript.

\section*{Conflicts of interest}
There are no conflicts to declare.

\section*{Acknowledgements}
P.M.P. and A.S. are grateful to Ulrike Diebold, Giada Franceschi, Florian Mittendorfer, and Andrea Conti for sharing some preliminary results of their ongoing work on this topic.
P.M.P. thanks Michiel Sprik for stimulating discussions about this work during his visit to Princeton University.
This work was conducted within the center Chemistry in Solution and at Interfaces funded by the USA Department of Energy under Award DE-SC0019394.
Simulations reported here were substantially performed using the Princeton Research Computing resources at Princeton University which is consortium of groups including the Princeton Institute for Computational Science and Engineering and the Princeton University Office of Information Technology’s
Research Computing department.
This research used resources of the Oak Ridge Leadership Computing Facility at the Oak Ridge National Laboratory, which is supported by the Office of Science of the U.S. Department of Energy under Contract No. DE-AC05-00OR22725.

\providecommand*{\mcitethebibliography}{\thebibliography}
\csname @ifundefined\endcsname{endmcitethebibliography}
{\let\endmcitethebibliography\endthebibliography}{}

\end{document}